\documentclass[11pt]{article}
\usepackage{geometry}
\usepackage{amsmath}
\usepackage{amsfonts}
\usepackage{amssymb}
\usepackage{graphicx}
\usepackage{algorithmic}
\usepackage{algorithm}
\usepackage{tikz}
\usepackage{makecell,multirow,todonotes,enumitem}
\usepackage{array}

\newcommand{\qed}{\hfill\hbox{\rule{6pt}{6pt}} \medskip}
\newcommand{\proof}{\noindent {\bf Proof}.\ }

\newcommand{\cB}{\mathcal{B}}
\newcommand{\cC}{\mathcal{C}}

\newcommand{\cH}{\mathcal{H}}

\def\CC{\mathbb{C}}

\def\imp{\to}

\newtheorem{theorem}{Theorem}[section]
\newtheorem{lemma}[theorem]{Lemma}

\newtheorem{corollary}[theorem]{Corollary}

\begin{document}

%\title{Approximation Algorithms \\ for Implication System Minimization}
\title{A Polynomial Time Approximation for the Minimization \\ of Implication Systems and Horn CNFs}
\date{\vspace{-5ex}}
\date{}  % Toggle commenting to test

\author{
Endre Boros\thanks{MSIS Department and RUTCOR, Rutgers University, 100 Rockafellar Road, Piscataway, NJ 08854, USA; (endre.boros@rutgers.edu)}
\and
Ond\v{r}ej \v{C}epek\thanks{Department of Theoretical Informatics, Charles University, Malostransk\'{e} n\'{a}m. 25, 118 00 Praha 1, Czech Republic; (ondrej.cepek@mff.cuni.cz)}
\and
Kazuhisa Makino\thanks{Research Institute for Mathematical Sciences (RIMS) Kyoto University, Kyoto 606-8502, Japan; (makino@kurims.kyoto-u.ac.jp)}
}

\maketitle
\thispagestyle{empty}

\begin{abstract}

An implication system is a collection of rules of the form $B \imp H$, where the set $B$ of Boolean variables is the {\em body} of the rule, while $H$ is the set of its {\em heads}. Such a rule is an aggregate representation of individual implications $B\imp h$, for $h\in H$ which correspond to the clauses of an equivalent  CNF representation of a Horn Boolean function. 
An implication system is {\em body-minimum}/{\em clause-minimum} if it has the minimum number of bodies/clauses among all logically equivalent implication systems. The corresponding minimization problems appear not only in the context of implication systems, but also as problems for closure systems, directed hypergraphs, representation of dependencies in relational databases, or in 
%(definite) 
Horn CNF minimization. 
While body-minimization is known to be solvable in polynomial time (and there are several algorithms in the literature that solve this task), clause-minimization is known to be NP-hard. In fact, body-minimization has a particular canonical solution, called the GD-basis of the underlying Horn function that also can be computed efficiently.
%In a Horn CNF context, this second minimization problem asks to find a logically equivalent Horn CNF with the minimum number of clauses. 

In this paper, we utilize a known connection between the two minimization problems stating that for every implication system there always exists a logically equivalent clause-minimum implication system which has the same bodies as its canonical GD-basis, see (B{\'e}rczi and B{\'e}rczi-Kov{\'a}cs, 2017), and (Adaricheva and Nation, 2014). 

We use this fact to design a polynomial time approximation algorithm which for any $\delta>0$ outputs a $(\frac{1}{4} + \delta)n$-factor approximation of the logically equivalent clause-minimum implication system on $n$ variables. This is an improvement from the best known algorithm so far that provides only an $(n-1)$-factor approximation, see (Hammer and Kogan, 1993).
\end{abstract}

\newpage

\section{Introduction}

We consider a finite set $V$ of Boolean variables that can take only values true (denoted by $1$) or false (denoted by $0$). For a given subset $B\subseteq V$ and variable $v\in V$, the implication $B\imp v$ is the logical expression that evaluates to $0$ if and only if all variables in $B$ have a value $1$ and $v$ equals $0$. We can read it as the statement ``if all variables in $B$ are $1$, then $v$ must also be $1$'' and we say that it is {\em satisfied} if it takes value $1$. The set $B$ is called the \emph{body} of the implication, while $v$ is its \emph{head}. For subsets $B,H \subseteq V$ we denote by $B\imp H$ the conjunction     
\[
\bigwedge_{h\in H} (B\imp h)
\]
and call it a {\em rule}. Note that $B\imp v$ is always true if $v\in B$, and hence we can assume w.l.o.g. that $B\cap H=\emptyset$. To simplify our notation, we write $B\imp h$ if $H=\{h\}$, and $b\imp H$ if $B=\{b\}$.
An \emph{implication system} $\Phi=\{B_i\imp H_i\mid i\in [m] \}$, where $[m]=\{1, \ldots ,m\}$, is a collection of rules with $B_i,H_i\subseteq V$ for $i\in [m]$, where the bodies $B_i$, $i\in [m]$ are assumed to be pairwise distinct and each set of heads $H_i$, $i\in [m]$ is assumed to be nonempty and disjoint from $B_i$ (see e.g.~\cite{Wild1994} for more details about implication systems). A truth assignment $x \in  \{0,1\}^V$ is a satisfying assignment of $\Phi$ if it satisfies all implications of $\Phi$. A \emph{Boolean function} $f$ on variables from $V$ is a mapping $f : \{0,1\}^V \imp \{0,1\}$. Clearly, every implication system $\Phi$ represents a Boolean function $f_{\Phi}$ on variables from $V$ where $f_{\Phi}(x) = 1$ if and only if $x \in  \{0,1\}^V$ is a satisfying assignment of $\Phi$.
Two implication systems $\Phi$ and $\Psi$ are \emph{equivalent}, denoted by $\Phi \equiv \Psi$, if and only if $f_{\Phi} = f_{\Psi}$.     

For instance, given four variables $V=\{a,b,c,d\}$ we can consider the following set of implications: 
$\Phi=\{a\imp b,~b\imp a,~\{a,c\}\imp b,~ \{a,c\}\imp d\}$. This set can be written in a more concise equivalent form as an implication system 
\begin{equation}\label{eq-example}
\Phi' ~=~ \left\{a\imp b,~b\imp a,~\{a,c\}\imp \{b,d\}\right\}.
\end{equation}
Setting variables $a$ and $b$ to $1$, and $c$ and $d$ to $0$ we get a satisfying assignment of $\Phi'$, while setting $a$, $b$, and $c$ to $1$ and $d$ to $0$ is not satisfying. 
%We shall call such unsatisfying assignments \emph{falsifying}. 

Implication systems allow us to extend our knowledge about a partial satisfying assignment. For instance, assume that the variables of a subset $S\subseteq V$ are $1$ in a satisfying assignment of the implication system $\Phi$, given as above. If there exists an index $1\leq i\leq m$ such that $B_i\subseteq S$ and $H_i\not\subseteq S$, then we can conclude that the variables in $H_i\setminus S$ must also be $1$, and hence we can extend $S$ to $S\cup H_i$. Repeating this as many times as possible, we arrive to a subset $S'\supseteq S$ such that for all indices $i\in [m]$ we have either $B_i\not\subseteq S'$ or $H_i\subseteq S'$. Such a subset $S'$ is called a \emph{closed} set with respect to $\Phi$. The \emph{closure} $\CC_\Phi(S)$ of an arbitrary subset $S\subseteq V$ is a minimal closed set with respect to $\Phi$ that contains $S$. Note that the closure of a set can be efficiently derived by the above simple procedure, which we call {\em forward chaining} following the standard terminology. Furthermore, the closure of a set $S$ depends only on the underlying Boolean function, in other words,  $\Phi \equiv \Psi$ if and only if $\CC_\Phi(S) = \CC_\Psi(S)$ for all $S\subseteq V$ (see e.g.~\cite{BB79,HK93}). In the sequel we call $\CC_\Phi$ the \emph{closure operator} of $\Phi$. Furthermore, we denote by $\cC(\Phi)$ the family of closed sets with respect to the implication system $\Phi$. Note again that  $\Phi\equiv \Psi$ if and only if  $\cC(\Phi)=\cC(\Psi)$. 
Note also that closed sets are in a one-to-one correspondence with satisfying assignments. Namely, a subset $S\subseteq V$ is a closed set with respect to $\Phi$ if and only if assigning $1$ to variables in $S$ and $0$ to variables not in $S$ is a satisfying assignment of $\Phi$.    
%Sets that are not closed are called \emph{open} with respect to $\Phi$.

By definition of the closure we have $B_i \cup H_i \subseteq \CC_\Phi(B_i)$ for $i\in [m]$, where some or even all of the containment relations may be strict.
%(and we shall see later that implication systems for which all these containment relations hold as equalities are called right-saturated). 
For instance, we can represent our small example above also as the equivalent implication system
\[
\Phi'' ~=~ \left\{a\imp b,~b\imp a,~\{a,c\}\imp d,~ \{b,c\}\imp d\right\},
\]
where e.g., $\CC_{\Phi'}(\{a,c\})  ~=~ \CC_{\Phi''}(\{a,c\}) ~=~ \{a,b,c,d\}$ so the containment is strict in $\Phi''$ for the body $\{a,c\}$, while it is not strict in $\Phi'$. Furthermore, we have $\{a,b\}\in \cC(\Phi')$, while $\{a,b,c\}\not\in \cC(\Phi')$. 

Let us recall that implication systems arise naturally in many different applications, and in many different contexts, resulting in different looking representations and terminology. Implication systems are used to describe the structure of relational databases \cite{A74,BB79,Maier80}, are equivalent to the family of definite Horn Boolean functions \cite{F82}, and can equivalently be viewed as directed hypergraphs \cite{ADS83,ADS86}. They arise in formal concept analysis \cite{GD86}, in closure systems \cite{Wild1994,AN14}, in machine learning \cite{AB09,AB11}, in the theory of Boolean functions (Horn theory) \cite{H51,McK43,BCKK10,CH11}, etc.
The family of Boolean functions that can be represented by implication systems is known as \emph{definite Horn functions} (in some literature called \emph{pure} Horn functions). The above cited forward chaining procedure that derives the closure of a set is known to run in polynomial time in the size of the input implication system (see e.g.~\cite{BB79,HK93}).

Our small examples above (implication systems $\Phi'$ and $\Phi''$) show that a Boolean function (a closure operator) may be represented by different logically equivalent implication systems. Since the computational time of various tasks, such as computing the closure of a set or answering a query in an expert system, depends on the implication system we use as an input, it is a frequently arising and important problem to find the shortest representation of a given implication system. 
%A natural measure of the size of an implication system is the number of rules in the description, which is $3$ in $\Phi'$ and $4$ in $\Phi''$, or equivalently, the number of distinct bodies used. This brings us to the main problem we consider in this paper, the {\em body-minimization problem:} Given an implication system $\Phi$, find a logically equivalent system $\Phi'$ that has the smallest possible number of distinct bodies. 

Since implication systems are used in the different contexts under different formalisms (e.g. directed hyperarcs, definite horn CNFs) there are numerous natural measures of ``size'' of an input and hence also numerous notions of ``minimality''. Six different measures and the corresponding concepts of minimality were considered for an implication system $\Phi = \{B_i \imp H_i \mid i \in [m]\}$ in~\cite{ADS86,CH11}: 
\begin{itemize}
    \item (B) number of bodies: $m$, 
    \item (BA) body area: $\sum_{i=1}^m |B_i|$,
    \item (C) number of clauses: $\sum_{i=1}^m |H_i|$,
    \item (TA) total area: $\sum_{i=1}^m (|B_i|+|H_i|)$,
    \item (BC) number of bodies and clauses: $\sum_{i=1}^m (|H_i|+1)$, and
    \item (L) number of literals: $\sum_{i=1}^m (|B_i|+1)\cdot|H_i|$.
\end{itemize}
With a slight abuse of notations we shall use (B), (BA), (TA), (C), (BC) and (L) to denote both the measures and the corresponding minimization problems. 

%(SM) minimizes the number of adjacency lists for a hypergraph or equivalently the number of rules for an implicational system, and hence it is a very natural measure for the size of the representation in these two contexts. On the other hand, it is not very natural in the CNF context and thus it is rarely used there. (CM) minimizes the number of hyperarcs or equivalently the number of clauses in the CNF, so it is one of the two common measures for CNFs.  (SHM) and (SAM) are slightly strange mixed size measures and do not appear in many other papers. (TA) is simply the total size of the adjacency list representation of a directed hyperarc and similarly a total size of an implicational system. It should be noted that the second common measure for CNFs, namely the total number of literals (sum of clause lengths) - let us denote it by (L) - is missing in the above list (in the hypergraph context (L) is the total size of the representation if all hyperarcs are listed individually). In our example the representations are already minimal and the sizes are as follows: (SM) three, (CM) four, (SHM) seven, (SAM) four, (TA) eight, and (L) ten. 

The only one of these six minimization problems for which a polynomial time procedure exists to derive a minimum representation is (B). The first such algorithm appeared in database theory literature~\cite{Maier80}. Different algorithms for the same task were then independently discovered in hypergraph theory~\cite{ADS86}, and in the theory of closure systems~\cite{GD86}.

For the remaining five measures it is NP-hard to find the shortest representation. There is an extensive literature on the intractability results in various contexts for these minimization problems~\cite{Maier80,ADS86,HK93}. It was shown that (C) and (L) stay NP-hard even when the inputs are limited to bodies of size at most two~\cite{BCK13}, and the same result extends to the remaining three measures. Note that if all bodies are of size one then the above problems become equivalent with the transitive reduction of directed graphs, which is tractable \cite{AGU72}. It should be noted that there exist several other tractable subclasses, see ~\cite{HK95,BCKK09}. For the subclass of key-Horn implication systems, it was shown that (BA) becomes tractable while the remaining four problems stay NP-hard~\cite{K17} but all admit a logarithmic factor approximation~\cite{BBCKM22}. There are also few older heuristic minimization approaches for (C) and (L)~\cite{BCK98}. 

It was shown that (C) and (L) are not only hard to solve exactly but even hard to approximate. More precisely, \cite{turan} shows that these problems are inapproximable within a factor $2^{\log^{1-\varepsilon}(n)}$ assuming $NP\subsetneq DTIME(n^{polylog(n)})$, where $n$ denotes the number of variables. In addition, \cite{BG14} shows that they are inapproximable within a factor $2^{\log^{1-o(1)}n}$ assuming $P\subsetneq NP$. It is not difficult to see that the same proof extends to (BC) and (TA) as well. On the positive side, (C), (BC), (BA), and (TA) admit $(n-1)$-approximations and (L) has an $\binom{n}{2}$-approximation \cite{HK93}. To the best of our knowledge, no better approximations are known.

In this paper, we study problems (B) and (C). As already mentioned above, there are several polynomial time minimization algorithms for (B) in the literature and they can be divided into two types. The ones from database theory~\cite{Maier80} and hypergraphs~\cite{ADS83,ADS86} use a local exchange lemma to decrease the number of bodies, utilizing an equivalence relation on the set of bodies. The algorithms from the closure systems literature~\cite{GD86,AB09,AB11} perform a sequence of steps, called {\em right-saturation, left-saturation, and redundancy elimination} to transform the input into an equivalent body-minimum representation. The latter approach produces a {\em unique} implication system representing the underlying Boolean function, which is called the \emph{GD-basis} of the input. This name is derived from the initial letters of the authors' names of~\cite{GD86}. 

In this paper we provide a first non-trivial polynomial time approximation algorithm for clause-minimization (problem (C)). Our result has three main components. First, in Section 3, we recall that there always exists a solution to problem (C) that has the same bodies as the unique GD-basis (which is computable in polynomial time). This result from~\cite{AN14,BB17} immediately implies that the GD-basis constitutes a polynomial time $(n-1)$-factor approximation for the clause minimization problem, matching the bound from~\cite{HK93}. Next, in Section 4, we show that the implication subsystem corresponding to the set of bodies with the same closure can be replaced, without destroying the logical equivalence with the underlying Horn function, by any other form using the same bodies and guaranteeing the same closure for those bodies. Finally, in Section 5 we provide a constructive approach to build such a replacement of "small" size, yielding 
a polynomial time approximation algorithm which for any $\delta>0$ outputs a $(\frac{1}{4} + \delta)n$-factor approximation for problem (C).

\section{Preliminaries}
\label{sec-def}
In this section, 
we introduce more notation to extend the definitions from the previous section, and show complexity results for several basic problems for implication systems.

For an implication system $\Phi=\{B_i\imp H_i\mid i\in [m]\}$ let us denote by $\cB(\Phi)=\{B_i\mid i\in [m]\}$ the set of bodies appearing in $\Phi$, 
%and by $|\Phi|=|\cB(\Phi)|$ the number of rules in $\Phi$. 
by $\|\Phi\| = \sum_{i=1}^m (|B_i|+|H_i|)$ the size of $\Phi$, and by $cnf(\Phi)=\sum_{i=1}^m |H_i|$ the number of individual implications (clauses) in $\Phi$. 

%To state our main results, we need to recall a few more definitions and prove a number of technical claims. 
An implication is a \emph{consequence} of an implication system if it is satisfied by every satisfying assignment of the system. The following lemma, cited from the literature without proof, provides a condition for checking this property. 

\begin{lemma}[\cite{HK93}]\label{lem-fc}
Let $B \imp v$ be an implication and $\Phi$ an implication system. Then $B \imp v$ is a logical consequence of $\Phi$ if and only if $v \in \CC_{\Phi}(B)$. Furthermore, the closure $\CC_\Phi(S)$ can be computed in $O(\|\Phi\|)$ time for every set $S\subseteq V$.\qed
\end{lemma}

Recall, that  for  a given implication system $\Phi$ we denote by $\cC(\Phi)=\{X\subseteq V\mid \CC_\Phi(X)=X\}$ the set of its closed sets and we have $\Phi\equiv \Psi$ exactly when $\cC(\Phi)=\cC(\Psi)$. 
 
\begin{lemma}\label{lem-equivalence-implications}
For two implication systems $\Phi=\{A_i\imp G_i\mid i\in [p]\}$ and $\Psi=\{B_j\imp H_j\mid j\in [q]\}$ we have $\Phi\equiv \Psi$ if and only if $\CC_\Phi(B_j)\supseteq H_j$ for all $j\in [q]$ and $\CC_\Psi(A_i)\supseteq G_i$ for all $i\in [p]$. All these conditions can be tested in $O(p\|\Psi\| + q\|\Phi\|)$ time.  
\end{lemma}
\proof
Recall that $X\in \cC(\Phi)$ iff $\CC_\Phi(X)=X$, and analogously, $Y\in \cC(\Psi)$ iff $\CC_\Psi(Y)=Y$. Consequently, the first part of the statement follows by the above definitions and the definition of equivalence.
The complexity claim follows by Lemma \ref{lem-fc} since the above inclusions can be tested in $O(\|\Psi\|)$ time for every rule in $\Phi$ and in $O(\|\Phi\|)$ time for every rule in $\Psi$. 
\qed

\bigskip

%\section{GD-basis and clause minimization}
\section{Functional properties of implication subsystems}

For an implication system $\Phi=\{B_i\imp H_i\mid i\in [m]\}$ and subset $I\subseteq [m]$ we say that $\Psi=\{B_i\imp H_i\mid i\in I\}$ is an \emph{implication subsystem} of $\Phi$. We will write $\Phi\setminus \Psi=\{B_i\imp H_i\mid i\in [m]\setminus I\}$ and $\Phi =(\Phi\setminus \Psi)\cup \Psi$. In other words, we view implication systems as sets of rules.

To an implication system $\Phi = \{ B_i \imp H_i \mid i \in [m] \}$ and subset $S \subseteq V$ we associate the following implication subsystem
\begin{eqnarray*}
\Phi_{\subseteq S} &=& \{B_i \imp H_i \mid i \in [m], \CC_\Phi(B_i) \subseteq \CC_\Phi(S) \}\\
\Phi_{\subsetneq S} &=& \{B_i \imp H_i \mid i \in [m], \CC_\Phi(B_i) \subsetneq \CC_\Phi(S) \}\\
\Phi_{=S} &=& \{B_i\imp H_i\mid i \in [m], \CC_\Phi(B_i) = \CC_\Phi(S)\}\\
\Phi_{\neq S} &=& \{B_i\imp H_i\mid i\in [m], \CC_\Phi(B_i)\neq \CC_\Phi(S)\}.
\end{eqnarray*}
These subsystems have important functional properties, which we show in the following lemmas.

\begin{lemma}\label{lem-subfunction-subsetneq}
    If $\Phi\equiv \Psi$ are equivalent implication systems, then for all subsets $S\subseteq V$ we have $\Phi_{\subsetneq S}\equiv \Psi_{\subsetneq S}$.
\end{lemma}
\proof
We are going to prove the claim by Lemma \ref{lem-equivalence-implications}. Let us consider an arbitrary subset $S\subseteq V$. Since in the claim the role of $\Phi$ and $\Psi$ are symmetric, it is enough to show that for all rules $B\imp H$ of $\Phi_{\subsetneq S}$ we have $\CC_{\Psi_{\subsetneq S}}(B)\supseteq H$. By the definition of $\Phi_{\subsetneq S}$ we have $\CC_\Phi(B)\subsetneq \CC_\Phi(S)$ implying that for all rules $B_i\imp H_i$ used in deriving $\CC_\Phi(B)$ we have $\CC_\Phi(B_i)\subsetneq \CC_\Phi(S)$ implying $\CC_{\Phi_{\subsetneq S}}(B)=\CC_{\Phi}(B)$. Thus we have $\CC_\Phi(S)\supsetneq \CC_{\Phi_{\subsetneq S}}(B)=\CC_{\Phi}(B)=\CC_\Psi(B)$, where the last equality follows by the equivalence of $\Phi$ and $\Psi$. Consequently, all rules of $\Psi$ used to derive $\CC_\Psi(B)$ are also rules of $\Psi_{\subsetneq S}$ implying $\CC_{\Phi_{\subsetneq S}}(B)=\CC_{\Phi}(B)=\CC_\Psi(B)=\CC_{\Psi_{\subsetneq S}}(B)$. Since $H\subseteq \CC_{\Phi_{\subsetneq S}}(B)$ by the choice of the rule $B\imp H$, the claim follows.
\qed

\begin{lemma}\label{lem-subfunction-subseteq}
    For two implication systems, $\Phi$ and $\Psi$ we have $\Phi\equiv \Psi$ if and only if $\Phi_{\subseteq S}\equiv \Psi_{\subseteq S}$ holds for all subsets $S\subseteq V$. 
\end{lemma}
\proof
The ``if'' part of the claim follows by the simple facts that $\Phi=\Phi_{\subseteq V}$ and $\Psi=\Psi_{\subseteq V}$. The ``only if'' part follows by an analogous proof as in the previous lemma, replacing all $\subsetneq$ relations in that proof by $\subseteq$. 
\qed

\begin{lemma}\label{lem-bullet}
    For an implication system $\Phi$ and a subset $S\subseteq V$ we have $\CC_{\Phi_{\neq S}}(T)=\CC_{\Phi_{\subsetneq S}}(T)$ for all subsets $T\subseteq \CC_\Phi(S)$.
\end{lemma}
\proof
Note that $\Phi_{\subsetneq S}$ is an implication subsystem of $\Phi_{\neq S}$ which is an implication subsystem of $\Phi$, implying by the definition of the closure operator that 
\begin{equation}\label{e-E1}
    \CC_{\Phi_{\subsetneq S}}(T) \subseteq \CC_{\Phi_{\neq S}}(T) \subseteq \CC_\Phi(T)
\end{equation}
for all subsets $T\subseteq V$. Since $T\subseteq \CC_{\Phi}(S)$ we also have 
\[
\CC_\Phi(T) \subseteq \CC_{\Phi}(S)
\]
by the properties of a closure operator.
Consequently for all implications $B_i\imp H_i$ used in deriving $\CC_{\Phi_{\neq S}}(T)$ we have $B_i\subseteq \CC_{\Phi_{\neq S}}(T)$ and consequently by the above relations and the properties of the closure operation we have $\CC_\Phi(B_i)\subseteq \CC_\Phi(S)$. The nonequality $\CC_\Phi(B_i)\neq \CC_\Phi(S)$ follows by the definition of $\Phi_{\neq S}$, implying 
$\CC_\Phi(B_i)\subsetneq \CC_\Phi(S)$, from which it follows that we have equality in the first inequality of \eqref{e-E1}.
\qed

\begin{lemma}\label{lem-gradual-leftsaturation}
    For an implication system $\Phi = \{ B_i \imp H_i \mid i \in [m] \}$ and index $j\in [m]$ we define $\Psi=(\Phi\setminus\{B_j\imp H_j\})\cup \{\CC_{\Phi_{\neq B_j}}(B_j)\imp H_j\}$. Then we have $\Phi\equiv \Psi$. 
\end{lemma}
\proof
We can apply Lemma \ref{lem-equivalence-implications} which in this special case implies that we have $\Phi\equiv \Psi$ if and only if $\CC_\Psi(B_j)\supseteq H_j$. Note next that $\Phi_{\neq B_j}$ is an implication subsystem of $\Phi\setminus\{B_j\imp H_j\}$ and thus it is an implication subsytem of $\Psi$. Consequently, we have $\CC_\Psi(B_j)\supseteq \CC_{\Phi_{\neq B_j}}(B_j)$ 
from which $\CC_\Psi(B_j)\supseteq H_j$ follows by the definition of $\Psi$.
\qed

To an implication system $\Phi = \{ B_i \imp H_i \mid i \in [m] \}$  we associte 
\[
\Phi^\bullet ~=~ \{ \CC_{\Phi_{\neq B_i}}(B_i)\imp H_i \mid i\in [m] \}
\]
and call it the \emph{left-saturation} of $\Phi$. 

\begin{corollary}\label{cor-leftsaturationn}
    For every implication system $\Phi$ we have $\Phi\equiv \Phi^\bullet$.
\end{corollary}
\proof
Follows by the sequential application of Lemma \ref{lem-gradual-leftsaturation}.
\qed

To an implication system $\Phi = \{ B_i \imp H_i \mid i \in [m] \}$  we associte 
\[
\Phi^* ~=~ \{ B_i\imp \CC_\Phi(B_i) \mid i\in [m] \}
\]
and call it the \emph{right-saturation} of $\Phi$. 

\begin{lemma}\label{lem-rigthsaturation}
    For every implication system $\Phi$ we have $\Phi\equiv \Phi^*$.
\end{lemma}
\proof
Immediate by Lemma \ref{lem-equivalence-implications}.
\qed

Let us finally associate to an implication system $\Phi$ the implication system 
$\Phi^{\circledast}$ obtained from $(\Phi^\bullet)^*$ by eliminating the redundant rules. Let us point out that if $\Phi$ is an irredundant implication system, then its left saturation will also be irredundant. However, it may contain rules that become redundant after right saturation. 
More precisely, introducing $B_i^\bullet = \CC_{\Phi_{\neq B_i}}(B_i) = \CC_{\Phi_{\subsetneq B_i}}(B_i)$ (where the second equality holds by Lemma~\ref{lem-bullet}) we may have pairs of rules in $\Phi^\bullet$ such that
\begin{eqnarray}
    %H_i \subseteq B_i^\bullet && \text{for some } i \in [m] \label{eq-redundant-head}\\
    B_j^\bullet \subseteq B_i^\bullet &\text{ and }~~~ \CC_\Phi(B_i) = \CC_\Phi(B_j)& \text{for some } i \neq j \in [m]. \label{eq-redundant-body}
\end{eqnarray}
In this case, rule $i$ can be replaced by rule $j$ extended with the heads of rule $i$ without losing equivalence with the underlying function. For details and a full proof we refer the reader to~\cite{Maier80}.

\begin{theorem}[\cite{GD86,AB09,AB11}]\label{theorem-GD-basis}
    For an implication system $\Phi$ the associated implication system $\Phi^\circledast$ is a unique body-minimum representation of the underlying Horn function, called its GD-basis.
    \qed
\end{theorem}

Note that the GD-basis $\Phi^\circledast$ can be computed in $O(m\| \Phi \|)$ time from any implication system $\Phi$ representing the same underlying Horn function, see e.g., \cite{GD86,AB09,AB11}.

The next theorem plays an important role in our constructive approximation result, and thus, for completeness,  we include here a short proof, which utilizes the above structural results of implications subsystems, and is slightly different from the proofs in the literature.

\begin{theorem}[\cite{AN14,BB17}]\label{theorem-clause-GD}
Every implication system $\Phi$ has a clause-minimum representation $\Sigma$ such that $\cB(\Sigma)=\cB(\Phi^\circledast)$.
\end{theorem}
\proof
Let $\Psi=\{B_i\imp H_i\mid i\in [m]\}$ be an arbitrary clause-minimum representation of $\Phi$ and define $\Pi=\Psi^\bullet=\{B_i^\bullet\imp H_i\mid i\in [m]\}$. Note that $\Psi$ is an irredundant system, and thus $\Pi$ is also irredundant (see e.g., \cite{AB09,AB11}). Then by Corollary \ref{cor-leftsaturationn} we have $\Pi\equiv \Psi$. Furthermore, by the definition of the left-saturation, we have $cnf(\Pi)=cnf(\Psi)$. As we noted above, $\Pi$ may contain pairs of rules  as in \eqref{eq-redundant-body}. In such a case, we can replace $B_i^\bullet\imp H_i$ and $B_j^\bullet\imp H_j$ by $B_j^\bullet\imp H_i \cup H_j$ which by Lemma~\ref{lem-equivalence-implications} preserves the equivalence with the underlying function (see also~\cite{Maier80}). Since $\Psi$ is clause-minimum, we must have $H_i\cap H_j=\emptyset$, and thus this operation does not change the number of clauses. 
After eliminating all such pairs, we arrive at an implication system $\Sigma$ for which we have $\Sigma\equiv \Psi$, $cnf(\Sigma)=cnf(\Psi)$ and $\cB(\Sigma)=\cB(\Psi^\circledast)$, where the last equality follows by the definition of the GD basis.
\qed

\begin{corollary}[\cite{AN14,BB17}]\label{c22}
Given an implication system $\Psi$ there exists an equivalent implication system $\Phi$ which is both clause-minimum and body-minimum. 
\end{corollary}
\proof
Follows directly by Theorem \ref{theorem-clause-GD}.
\qed

\section{Decompositions and substitutions}

In this section, we provide a number of results about substituting an implication subsystem by another implication system without losing equivalence with the underlying Horn function.

\begin{lemma}\label{lem-union-of-systems}
Assume $\Pi$, $\Phi$ and $\Psi$ are implications systems such that $\Pi=\Phi\cup \Psi$.
Then we have $\cC(\Pi)=\cC(\Phi)\cap \cC(\Psi)$. 
\end{lemma}
\proof
Follows directly from the definition of a closed set.
\qed

\begin{lemma}\label{lem-subfunctions}
Assume $\Phi=\Phi'\cup \Phi''$ and $\Phi''\equiv \Psi$. Then we have $\Phi'\cup \Psi \equiv \Phi$. 
\end{lemma}
\proof
By Lemma \ref{lem-union-of-systems} and our assumptions, we have $\cC(\Phi)=\cC(\Phi')\cap \cC(\Phi'')$ and $\cC(\Phi'')=\cC(\Psi)$ from which $\cC(\Phi')\cap \cC(\Psi)=\cC(\Phi)$ follows, proving the equivalence as claimed.
\qed

\bigskip

Lemma \ref{lem-subfunctions} shows that an implication subsystem can be replaced by an equivalent expression, preserving equivalence of the implication system. The next lemma shows that certain subsystems can be switched to another expression preserving equivalence of the implication system, even if the switched expressions themselves are not necessarily equivalent.

\begin{lemma}\label{lem-=switch}
    Assume $\Phi\equiv \Psi$ and $S\subseteq V$. Then for 
    \[
    \Sigma=\Phi_{\subsetneq S} \cup \Psi_{=S} \cup (\Phi\setminus \Phi_{\subseteq S})
    \]
    we have $\Sigma \equiv \Phi$.
\end{lemma}
\proof
Let us consider $\Pi=\Psi_{\subseteq S}\cup (\Phi\setminus \Phi_{\subseteq S})$. By Lemma \ref{lem-subfunction-subseteq} we have $\Pi\equiv \Phi$. Then we have $\Sigma=\Phi_{\subsetneq S}\cup (\Pi\setminus \Pi_{\subsetneq S})$, and thus $\Sigma\equiv \Pi\equiv \Phi$ follows by Lemma \ref{lem-subfunction-subsetneq}.
\qed

The next lemma proves the interesting property that a clause minimum implication system is also locally minimum for certain subsytems.

\begin{lemma}\label{lem-cnfmin-subsystem}
Given an implication system $\Phi$, consider a clause-minimum $\Psi\equiv \Phi$. Then for every $S\subseteq V$ we have 
\[
cnf(\Psi_{=S}) = \min_{\Pi\equiv \Psi} cnf(\Pi_{=S})= \min_{\Pi\equiv \Psi_{\subseteq S}} cnf(\Pi_{=S}).
\]
\end{lemma}
\proof
Assume for a contradiction that there exists a $\Pi\equiv \Phi$ such that $cnf(\Pi_{=S})<cnf(\Psi_{=S})$ for some $S\subseteq V$. Then by Lemma \ref{lem-=switch} the implication system
\[
\Sigma~=~ \Psi_{\subsetneq S} \cup \Pi_{=S} \cup (\Psi\setminus \Psi_{\subseteq S})
\]
is equivalent with $\Psi$ and has 
\[
cnf(\Sigma)=cnf(\Psi_{\subsetneq S})+cnf(\Pi_{=S})+cnf(\Psi\setminus \Psi_{\subseteq S}) <cnf(\Psi)
\]
contradicting the clause-minimality of $\Psi$. 
This contradiction proves the first equality. The second equality follows by Lemmas \ref{lem-subfunction-subseteq} and \ref{lem-subfunctions}.
\qed

Finally, the next lemma proves an exchange property which is crucial for the approximation algorithm in the next section.

\begin{lemma}\label{lem-equivclass-substitution}
    Let us consider an implication system $\Phi$, a body $B\in \cB(\Phi)$ and another implication system $\Psi$ such that $\cB(\Psi)=\cB(\Phi_{=B})$ and $\CC_\Psi(B')=\CC_\Phi(B')$ holds for every $B'\in \cB(\Phi_{=B})$. Then
    \[
    \Pi = \Phi_{\subsetneq B}\cup \Psi \cup (\Phi\setminus \Phi_{\subseteq B})
    \]
    is equivalent to $\Phi$. 
\end{lemma}

\proof
Let us note that $\Phi^*\equiv \Phi$ by Lemma \ref{lem-rigthsaturation}, and we have $\Psi\equiv \Phi^*_{=B}$ by Lemma \ref{lem-equivalence-implications}. Thus $\Phi\equiv \Pi=\Phi^*_{\subsetneq B} \cup \Psi \cup (\Phi^*\setminus \Phi^*_{\subseteq B})$ by Lemma \ref{lem-subfunctions}, and therefore the claim follows by Lemma \ref{lem-=switch} applied to $\Phi$ and $\Pi$. 
\qed

\section{Approximation for clause minimization}

In this section, we present an approximation algorithm for the clause minimization problem. By Theorem~\ref{theorem-clause-GD} the clause minimization problem for an implication system $\Phi = \{B_i \imp H_i \mid i \in [m]\}$ is equivalent to finding a clause-minimum implication system, logically equivalent to $\Phi$, the bodies of which are the bodies of the GD-basis of $\Phi$. 
Since the GD-basis can be computed efficiently, in the rest of this section we assume that $\Phi=\Phi^{\circledast}$ is the input to the approximation algorithm. This allows a constructive approach which iteratively replaces subsystems in a way that preserves logical equivalence of the entire implication system. In particular, Lemmas \ref{lem-=switch}, \ref{lem-cnfmin-subsystem} and \ref{lem-equivclass-substitution} will be instrumental in our arguments. 

%Thus, according to Theorem~\ref{theorem-clause-GD}, our task is to find subsets $H_i^* \subseteq \CC_\Phi(B_i)$, $i \in [m]$ such that $\Psi = \{B_i \imp H_i^* \mid i \in [m]\} \equiv \Phi$ and $cnf(\Psi)$ is minimum. 

To an implication system $\Phi=\Phi^{\circledast}$ let us associate the family of body closures:
\[
\cH(\Phi) ~=~ \{ \CC_{\Phi}(B) \mid B\in \cB(\Phi) \}.
\]
For $H \in \cH(\Phi)$ we define
\[
\Phi_H = \{B \imp H\setminus B \mid B\in \cB(\Phi), ~\CC_\Phi(B)=H\}.
\]
Using this notation, $\Phi$ can be decomposed into subsystems 
\[\Phi = \bigcup_{H \in \cH(\Phi)}\Phi_H. 
\]
%Note that since we assume that $\Phi$ is the GD-basis of the underlying function, $\CC_\Phi(H)=H$ holds for every $H\in \cH(\Phi)$.

In our approximation algorithm, we apply two different constructions to replace each subsystem $\Phi_H$ by another implication system $\Psi_H$ such that $\Phi \equiv (\Phi \setminus \Phi_H) \cup \Psi_H$. 
Let us also introduce
\[
OPT_H ~=~ \min_{\Pi\equiv \Phi} cnf(\Pi_{=H}) ~~~\text{ and }~~~OPT=\min_{\Pi\equiv \Phi} cnf(\Pi)=\sum_{H\in \cH(\Phi)} OPT_H. 
\]
If $\Pi$ is an arbitrary clause-minimum implication system of $\Phi$, then we have $cnf(\Pi)=OPT$ and $cnf(\Pi_{=H})=OPT_H$ for all $H\in \cH(\Phi)$ by Lemma \ref{lem-cnfmin-subsystem}, which also justifies the last equality in the formula above.

\subsection*{Procedure 1}

Let us assume that we focus on such a  subsystem $\Phi_H$ for some $H\in \cH(\Phi)$, as defined above. Assume that 
$\cB(\Phi_H)=\{B_i \mid i \in [p]\}$, $p\geq 2$ is the family of bodies which are of course subsets of $H$. 
Let $\alpha_{ij}=|B_i \setminus B_j|$ for any pair of distinct indices $i, j\in [p]$,  and define $\alpha^*  =\displaystyle\min_{i \in H} \sum_{\substack{j \in [p]} \\ j\neq i} \alpha_{ij}$. Our first procedure is based on the following lemma.

\begin{lemma}\label{lem-alphastar}
$\alpha^* \leq\frac{|H|p}{4}$. 
\end{lemma}
\proof
For any $S \subseteq [p]$, 
let $\displaystyle U_S=(\bigcap_{i \in S}B_i)\cap (\bigcap_{i \not\in S} (H\setminus B_i))$. Note that these sets form a partition of $H$.
We then have 
\begin{eqnarray}
p \alpha^* \ \leq \  \sum_{i, j \in [p] \atop i\neq j}\alpha_{ij}&=&\sum_{S \subseteq [p]}(p-|S|)|S||U_S|.  \label{eqa111a}
\end{eqnarray}
If $p$ is even,  
the right-hand side of (\ref{eqa111a}) is bounded by 
$\frac{p^2}{4}\sum_{S \subseteq [p]}|U_S|=  |H|p^2/4$,   %\label{eqa111}
%\end{eqnarray}
which implies the stated inequality.  

On the other hand, if $p$ is odd, 
the right-hand side of (\ref{eqa111a}) is bounded by 
$((p+1)(p-1)\sum_{S \subseteq [p]}|U_S|)/4  =  (p+1)(p-1)|H|/4\leq |H|p^2/4$,   %\label{eqa111}
%\end{eqnarray}
which again implies the claim.  
\qed

{\small
\begin{center}
\fbox{
\parbox{4.5in}{
\centerline{{\sc $1$st Approximate Clause-Minimum}}
\vskip 1ex
\begin{tabbing}
\noindent{\bf Input}:\hspace{0.1in}\=An implication system $\Phi_H=\{B_i\imp H\setminus B_i \mid i \in [p]\}$.\\[.15cm]
\noindent{\bf Step 1}: \> Compute an index $i^* \in [p]$ such that 
$\alpha^*  =\sum_{j \in [p], j\neq i^*} \alpha_{i^*j}$.\\[.05cm]
\noindent{\bf Step 2}: \> 
Output $\Psi_H= \{ B_i \imp B_{i^*}\setminus B_i\mid i \in [p]\setminus \{i^*\}\} \cup \{B_{i^*}\imp H \setminus B_{i^*}\}$
\end{tabbing}}}
\end{center}
}
\begin{lemma}
\label{lemma-bound1}
The above procedure computes in $O(p^2|H|)$ time an implication system $\Psi_H$ equivalent to $\Phi_H$ such that
\[
cnf(\Psi_H)\leq (\frac{1}{4}+\frac{1}{p})|H|\cdot p\leq (\frac{1}{4}+\frac{1}{p})|H|\cdot OPT_H. 
\]
\end{lemma}

\proof
By its definition, we have $\CC_{\Psi_H}(B_i)=H$ for all $i\in [p]$, consequently $\Psi_H\equiv \Phi_H$ follows by Lemma \ref{lem-equivalence-implications}.
Furthermore, by Lemma \ref{lem-alphastar} we have
\[
cnf(\Psi_H)=\alpha^* +|H\setminus B_{i^*}| \leq \frac{|H|p}{4}+|H|\leq (\frac{1}{4}+\frac{1}{p})|H|\cdot p
\]
as claimed. Note finally that $OPT_H\geq p$, since each body of $\Phi_H$ must have at least one head in every equivalent representation, implying the second inequality. 
\qed

\subsection*{Procedure 2}

Since the previous approach may not provide a good enough approximation ratio when $p$ is not large enough, we also consider another, much simpler exhaustive procedure.

{\small
\begin{center}
\fbox{
\parbox{4in}{
\centerline{{\sc $2$nd Approx Clause-Minimum}}
\vskip1ex
\begin{tabbing}
\noindent{\bf Input}:\hspace{0.1in}\=An implication system 
$\Phi_H=\{B_i\imp  H\setminus B_i \mid i \in [p]\}$ and a constant $c$, \\ \> ~~~~~where we assume that $|H| \geq c$.\\[.15cm]
\noindent{\bf Step 1}: \> Search exhaustively for hypergraphs $\{H_i\subseteq H \setminus B_i \mid i \in [p]\}$ such that \\[.1cm]
\> ~~~~~$\sum_i |H_i| \leq cp$ ~and~ $\Phi_{\subseteq H} \equiv \Phi_{\subsetneq H} \cup \{B_i\imp H_i\mid i\in [p]\}$\\[.1cm]
\noindent{\bf Step 2}: \> 
If such hypergraphs exist, choose one with the smallest $\sum_i |H_i|$ and \\[.1cm]
\> ~~~~~define $\Psi_H= \{ B_i \imp H_i \mid i  \in [p] \}$, otherwise, define $\Psi_H=\Phi_H$\\[.1cm]
\noindent{\bf Step 3:} \>  Output  $\Psi_H$. 
\end{tabbing}}}
\end{center}
}

\begin{lemma}\label{lem-bound2}
    The above procedure computes in $O((np)^{cp}p\|\Phi\|)$ time an implication system $\Psi_h$ on $n$ variables such that $\Phi_{\subseteq H} \equiv \Phi_{\subsetneq H}\cup \Psi_H$ and 
    \[
    cnf(\Psi_H) \leq \frac{|H|}{c} OPT_H. 
    \]
\end{lemma}
\proof
It is not difficult to see that 
Step 1 in the above procedure
can be performed by checking $O((np)^{cp})$ many cases. In each case we need to check the stated equivalence, which is equivalent by Lemma \ref{lem-equivalence-implications} to check for each $i\in [p]$ if the closure of $B_i$ by the constructed implication form is indeed $H$. For each case this takes $O(p\|\Phi\|)$ time. If there is a solution in Step 1, then by Lemma \ref{lem-cnfmin-subsystem} we have $cnf(\Psi_H)=OPT_H \leq \frac{|H|}{c} OPT_H$ by our assumption $|H| \geq c$. 

On the other hand, if no solution is found in Step 1,then we have $OPT_H > cp$ again by Lemma \ref{lem-cnfmin-subsystem}, and thus for the output $\Psi_H=\Phi_H$ we have $cnf(\Phi_H)\leq p|H|=\frac{|H|}{c}pc < \frac{|H|}{c} OPT_H$, proving the claim.
\qed

\subsection*{Polynomial time approximation algorithm}

Now we are ready to state our main approximation result for clause minimization.

\begin{theorem}
Given an implication system $\Sigma$ on $n$ variables and a constant $\delta>0$,  
we can compute in polynomial time an implication system $\Psi\equiv \Sigma$  such that 
\[
cnf(\Psi) ~\leq~ (\frac{1}{4}+ \delta)n \cdot \min_{\Pi\equiv \Sigma} cnf(\Pi) = (\frac{1}{4}+ \delta)n \cdot OPT.
\]
\end{theorem}

\proof
Let us assume $n \geq 8$ since otherwise we can compute the clause minimum form in constant time.

Assume $m=|\cB(\Sigma)|$ and
let $\Phi=\Sigma^{\circledast}$ be the GD-basis of $\Sigma$. By \cite{AB09,AB11} it can be computed in $O(m\|\Sigma\|)$ time from $\Sigma$. 

Let us consider, as above, the closures of the bodies of the GD-basis, and define
\begin{eqnarray*}
\cH_1(\Phi) &=&\{H \in \cH(\Phi) \mid |\cB(\Phi_{=H})| > 1/\delta\}\\
\cH_2(\Phi) &=&\{H \in \cH(\Phi) \mid |\cB(\Phi_{=H})| \leq 1/\delta\}.
\end{eqnarray*}

For $H\in \cH_1(\Phi)$ we compute $\Psi_H$ by the first procedure and obtain by Lemma \ref{lemma-bound1} that 
\[
\frac{cnf(\Psi_H)}{OPT_H} \leq \left(\frac14 + \frac{1}{|\cB(\Phi_{=H})|}\right)|H| < \left(\frac14+\delta\right)|H|.
\]

For $H\in \cH_2(\Phi)$, $|H| \geq 4$, we apply the second procedure with $c=4$ to obtain $\Psi_H$ for which by Lemma \ref{lem-bound2} we have 
\[
\frac{cnf(\Psi_H)}{OPT_H} \leq \frac{|H|}{4}.
\]

For $H\in \cH_2(\Phi)$ such that $|H| < 4$ we can choose $\Psi_H = \Phi_H$ because then we automatically have 
\[
\frac{cnf(\Psi_H)}{OPT_H} \leq 2 \leq \frac{n}{4}.
\]
by our assumption $n \geq 8$. 

Since $cnf(\Psi)=\sum_{H\in \cH(\Phi)} cnf(\Psi_H)$ and $OPT=\sum_{H\in \cH(\Phi)} OPT_H$ we get 
\[
\frac{cnf(\Psi)}{OPT} ~\leq~ \max_{H\in \cH(\Phi)} \frac{cnf(\Psi_H)}{OPT_H} \leq \left(\frac14+\delta\right) n
\]
proving the claimed inequality.

In our computations, the dominant part is the execution of Step 1 in the second procedure, which for $p \leq 1/\delta$ and $c=4$ requires $O(n^{\frac{4}{\delta}}\|\Phi\|)$ time.
\qed

\end{document}